\documentclass{article}

\usepackage[final]{neurips_2022_ml4ps}

\usepackage[utf8]{inputenc} 
\usepackage[T1]{fontenc}    
\usepackage{hyperref}       
\usepackage{url}            
\usepackage{booktabs}       
\usepackage{amsfonts}       
\usepackage{nicefrac}       
\usepackage{microtype}      
\usepackage{xcolor}         
\usepackage{microtype}
\usepackage{graphicx}
\usepackage{booktabs} 
\usepackage{physics}
\usepackage{newtxtext, newtxmath}
\usepackage{amsmath}
\usepackage[T1]{fontenc}
\usepackage{ae,aecompl}
\usepackage{bm}
\usepackage{hyperref}
\usepackage{url}
\usepackage{algorithm}
\usepackage{algpseudocode}
\usepackage{graphicx}
\usepackage{booktabs}
\usepackage{array}
\usepackage{booktabs}
\usepackage{caption}
\usepackage{graphicx}	
\usepackage{amsmath}	
\usepackage{subcaption}
\usepackage{hyperref}

\newcommand*\diff{\mathop{}\!\mathrm{d}}

\title{Modeling halo and central galaxy orientations on the SO(3) manifold with score-based generative models}

\author{%
 Yesukhei Jagvaral\thanks{yjagvara@andrew.cmu.edu}, \, Rachel Mandelbaum 
  \\
  McWilliams Center for Cosmology \\
  NSF AI Planning Institute for Data-Driven Discovery in Physics \\
  Department of Physics,\\
  Carnegie Mellon University\\
  Pittsburgh, PA 15213 USA \\
  \And
 Fran\c{c}ois Lanusse  \\
 AIM, CEA, CNRS \\
 Universit\'e Paris-Saclay \\
  Universit\'e Paris Cit\'e\\
  91191 Gif-sur-Yvette, France
}

\begin{document}

\maketitle

\begin{abstract}
Upcoming cosmological weak lensing surveys are expected to constrain cosmological parameters with unprecedented precision. In preparation for these surveys, large simulations with realistic galaxy populations are required to test and validate analysis pipelines.  However, these simulations are computationally very costly -- and at the volumes and resolutions demanded by upcoming cosmological surveys, they are computationally infeasible.
Here, we propose a Deep Generative Modeling approach to address the specific problem of emulating realistic 3D galaxy orientations in synthetic catalogs. For this purpose, we develop a novel Score-Based Diffusion Model specifically for the SO(3) manifold. The model accurately learns and reproduces correlated orientations of galaxies and dark matter halos that are statistically consistent with those of a reference high-resolution hydrodynamical simulation.  
\end{abstract}


\section{Introduction}
\label{submission}
 
Future wide-field  astronomical imaging surveys, such as the Vera C.\ Rubin Observatory Legacy
Survey of Space and Time\footnote{ \url{https://www.lsst.org/} }, Roman Space Telescope\footnote{ \url{https://roman.gsfc.nasa.gov/} } High Latitude Survey and Euclid\footnote{ \url{https://www.euclid-ec.org/} } will provide precise constraints on cosmological parameters   by imaging billions of galaxies. 
Deriving physical understanding from these data will require increasingly costly large-volume simulations with high  resolution
to test and validate analysis pipelines \citep{buzzard-1, buzzard-2,2019ApJS..245...26K} and to constrain cosmology via Simulation-Based Inference \citep[SBI;][]{des-lfi}. 

In this regard generative machine learning approaches represent an interesting avenue as they could serve as fast and robust emulators to greatly accelerate parts of the simulation pipelines. In particular, they could be used to populate realistic galaxies in large volume dark matter only simulations.  
Most machine learning methods in this line of research have been concerned with modeling scalar properties of galaxies, however in this work we are particularly interested in modeling the 3D orientations of galaxies and their host dark matter halos in simulations. These intrinsic orientations can indeed contaminate measurements of weak gravitational lensing in upcoming surveys and constitute a major source of systematic errors if not accounted for \citep{2015SSRv..193....1J}. 

Diffusion models are flexible in their domains that the datal ives, we want to jointly model various properties that live on various different spaces/manifolds

Currently, score-based diffusion models represent the state-of-the-art in generative tasks such as: image, audio and molecules generation.  \citep{hoogeboom22a}. 
Modeling distributions on the manifold of 3D rotations is however a non trivial task, and to address
this problem we develop a new type of score-based diffusion model specifically for the SO(3) manifold, by extending the Euclidean framework introduced in \cite{song2021}. We chose diffusion models due to their flexibility to model data that live on various different spaces (e.g. scalars and rotation matrices) compared normalizing flows and due to their stability compared to Generative Adversarial Networks. 
Based on these developments, we build a conditional generative model on SO(3) which allows us to sample from the posterior distribution of 3D orientations of galaxies and dark matter halo given information about their surrounding gravitational tidal field.




\section{Related Work}
Machine learning approaches have been adopted in astrophysics and cosmology in various contexts, including emulation methods, inference and forward modeling \citep{Dvorkin:2022pwo}. 
In particular, deep generative models have been implemented in the works of   \citet{graphgan} for generative modeling of correlated galaxy properties, such as shapes and orientations, with graph-based generative adversarial networks. 
Our work takes the next step to build generative models for various galaxy properties associated with galaxy and halo orientations (which are described by a non-Euclidean manifold) with score-based denoising diffusion models. 

\section{Score-Based Generative Model on SO(3)}
\label{sec:sgm}

Here we briefly outline our novel approach for modeling distributions on SO(3), heavily inspired by the diffusion framework developed in \cite{song2021}. The idea behind diffusion models is to introduce a noising process that perturbs the data distribution until it reaches a nearly pure noise distribution. Consider the following Stochastic Differential Equation (SDE) on the SO(3) manifold:
\begin{equation}
    \diff X = \mathbf{f}(X, t)  \diff t + g(t) \diff W, \label{eqn:forward_sde}
\end{equation}
where $W$ is a Brownian process on SO(3), $\mathbf{f}(\cdot \ , t): \text{SO(3)} \rightarrow T_{X}$SO(3) is a drift term, and $g(\cdot): \mathbb{R} \rightarrow \mathbb{R}$ is a diffusion term. Given samples $X(0) \sim p_\text{data}$ from an empirical data distribution $p_\text{data}$ at time $t=0$, the marginal distribution of samples $X(t)$ evolved under this SDE at a subsequent time $t > 0$ will be denoted $p_t$, and will converge for large $t=T$ towards a given predetermined distribution $p_T$ typically chosen to be easy to sample from. On SO(3), a natural choice for $p_T$ is $\mathcal{U}_{SO(3)}$, the uniform distribution on SO(3).

The key realization of \cite{song2021} is that under mild regularity conditions this noising process of the data process can be reversed, in particular through the following so-called probability flow Ordinary Differential Equation (ODE):
\begin{equation}
    \mathrm{d} X = [\mathbf{f}(X, t) - g(t)^2  \nabla \log p_t(X)] \mathrm{d} t .\label{eqn:probability_flow_ODE}
\end{equation}
\cite{de_bortoli_riemannian_2022} recently extended this result to compact Riemannian manifolds, which include in particular SO(3). 
This deterministic process is entirely defined as soon as the \textit{score function} $\nabla \log p_t(X) \in T_X \text{SO(3)}$ is known, and running this ODE backward in time from samples $X(T) \sim p_T$ down to $t=0$ will yield samples $X(0) \sim p_0 = p_\text{data}$. Training such a generative model will therefore boil down to estimating this score function with a neural network. 

While these results are direct analogs of Euclidean diffusion models \citep[as in][]{song2021}, implementing similar models on SO(3) brings practical difficulties: Unlike in the Euclidean case where the Gaussian is a closed-form solution of heat diffusion (a key element in Euclidean SGMs), there is no closed-form solution on general Riemannian manifolds. 
Our contribution is to propose solutions to these issues in order to implement efficient score-based diffusion models on SO(3).

On SO(3), although the exact heat kernel is only available as an infinite series \citep{nikolayev70}, it can be robustly approximated in practice either by truncating this series or by using a closed form expression \citep{Matthies80}, depending on the width of the kernel. It is used to define the so-called Isotropic Gaussian Distribution on SO(3),  $\mathcal{IG}_{\text{SO(3)}}(R, \epsilon)$ \cite{nikolayev70,Matthies80,leach22}, where $R \in \text{SO(3)}$ is a mean rotation matrix, and $\epsilon$ a scale parameter. $\mathcal{IG}_{\text{SO(3)}}$ enjoys tractable likelihood evaluation and sampling, and most importantly is closed under convolution. 

We can now define a noise kernel $p_\epsilon(X | X^\prime)=\mathcal{IG}_{\text{SO(3)}}(X ; X^\prime, \epsilon )$ which can be used to convolve the data distribution such that $ p_\epsilon(X) = \int_{SO(3)} p_\text{data}(X^\prime) p_\epsilon(X | X^\prime) \diff X^\prime $. For simplicity, we further make the following specific choice, for the diffusion SDE \autoref{eqn:forward_sde}: $\mathbf{f}(X, t) = 0$, $g(t)=\sqrt{\frac{\diff \epsilon(t)}{\diff t}}$ where $\epsilon(t)$ is a given noise schedule (e.g. $\epsilon(t)=t$). We then recover that convolving the data distribution with an $\mathcal{IG}_{\text{SO(3)}}$ of scale $\epsilon(t)$ corresponds to the marginal distribution of the SDE at time $t$: $p_{\epsilon(t)} = p_t$.


This noise kernel allows us to use on SO(3) the usual Denoising Score-Matching loss at no extra complexity compared to the Euclidean case. To learn the score function we introduce a neural score estimator $s_\theta(X, \epsilon) : \text{SO(3)}\times\mathbb{R}^{+ \star} \rightarrow \mathbb{R}^3$, which we train under the following loss:
\begin{equation}
    \mathcal{L}_{DSM} = \mathbb{E}_{p_\text{data}(X)} \mathbb{E}_{\epsilon \sim \mathcal{N}(0, \sigma_\epsilon^2)} \mathbb{E}_{p_{|\epsilon|}(\tilde{X} | X )} \left[ |\epsilon| \ \parallel   s_\theta(\tilde{X}, \epsilon) - \nabla \log p_{|\epsilon|}( \tilde{X} | X) \parallel_2^2 \right]
    \label{eqn:dsm}
\end{equation}
where we sample at training time random noise scales $\epsilon \sim \mathcal{N}(0, \sigma_\epsilon^2)$ similarly to \cite{song_improved_2020}. The minimum of this loss will be achieved for $s_\theta(X, \epsilon) = \nabla \log p_{\epsilon}(X)$. 

Once the score function is learned through $\mathcal{L}_{DSM}$, we can plug it in \autoref{eqn:probability_flow_ODE}, yielding the following sampling procedure given our specific choices for the SDE terms:
\begin{equation}
    X_T \sim \mathcal{U}_\text{SO(3)}  \qquad ; \qquad \diff X_t = -\frac{1}{2} \frac{\diff \epsilon(t)}{\diff t} s_\theta(X_t, \epsilon(t)) \diff t \;. \label{eqn:sampling_ode} 
\end{equation}
We solve this ODE down to $t=0$ to yield samples from the learned distribution. We illustrate this process in \autoref{tbl:toy_density}. Note that this is a manifold-valued ODE, which we solve using Runge-Kutta-Munthe-Kaas (RK-MK) algorithms for ODEs on Lie Groups (and we direct the interested reader to \cite{lie-methods} for a review). Finally, we note that this generative model can trivially be made conditional, by conditionning $s_\theta(X, t, y)$ on external information $y$ during training and sampling.

  \newcolumntype{M}[1]{>{\centering\arraybackslash}m{#1}}

    \begin{figure}[t]
        \centering
        \begin{tabular}{M{23mm}M{23mm}M{23mm}M{23mm}M{23mm} }
      \\
            $t=T$ & $t=3T/4$ & $t=T/2$ & $t=T/4$ & $t=0$  \\
            \includegraphics[width=26mm]{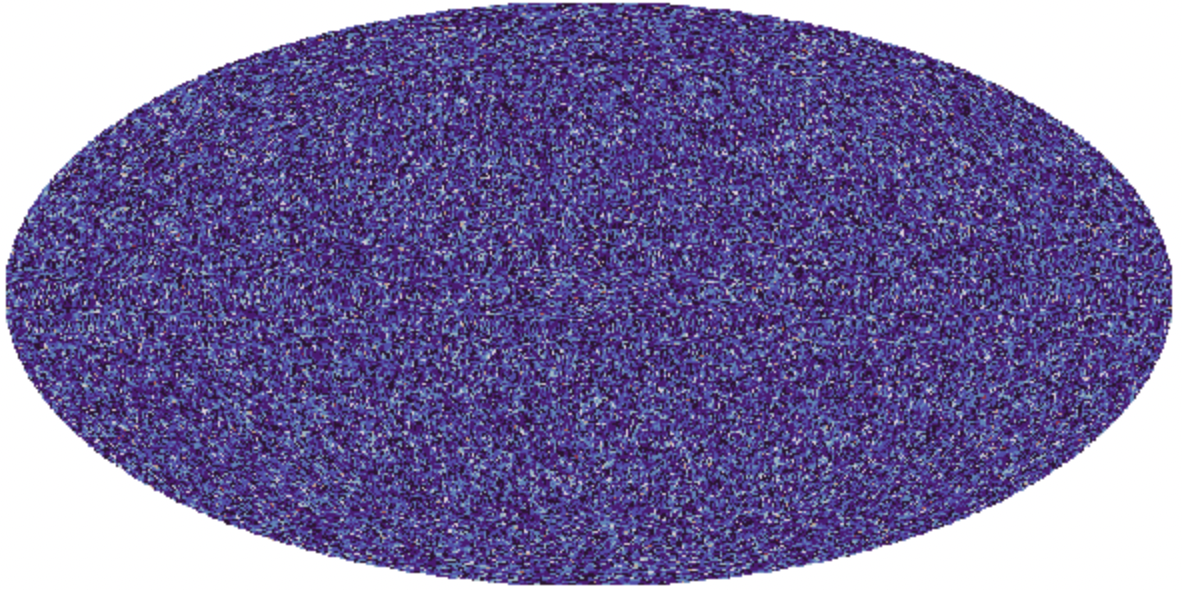} & \includegraphics[width=26mm]{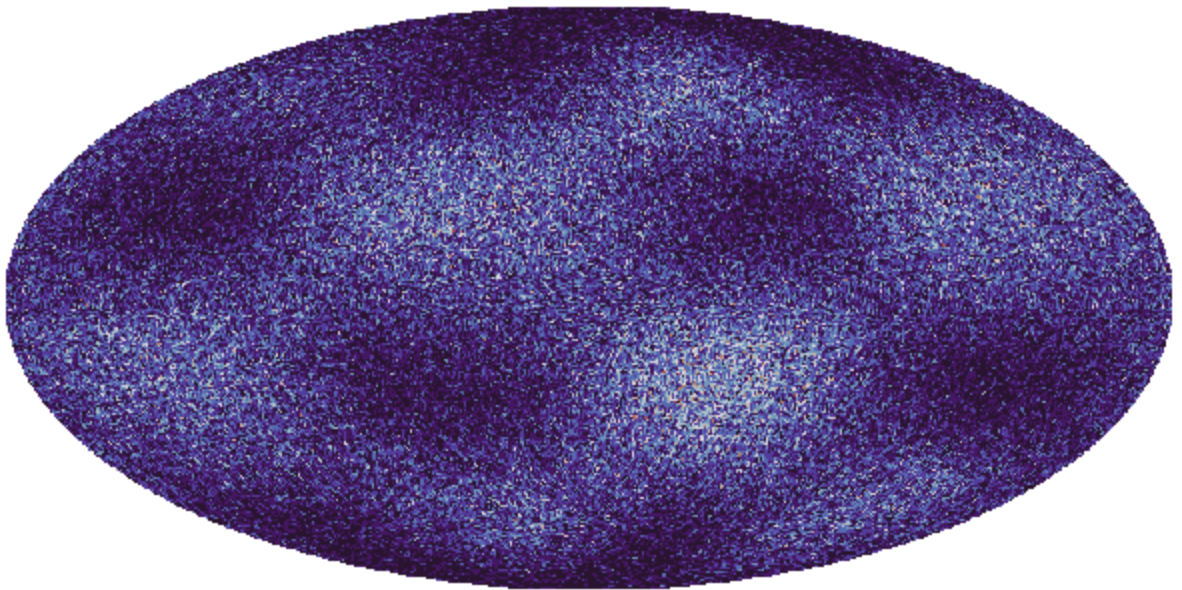} & \includegraphics[width=26mm]{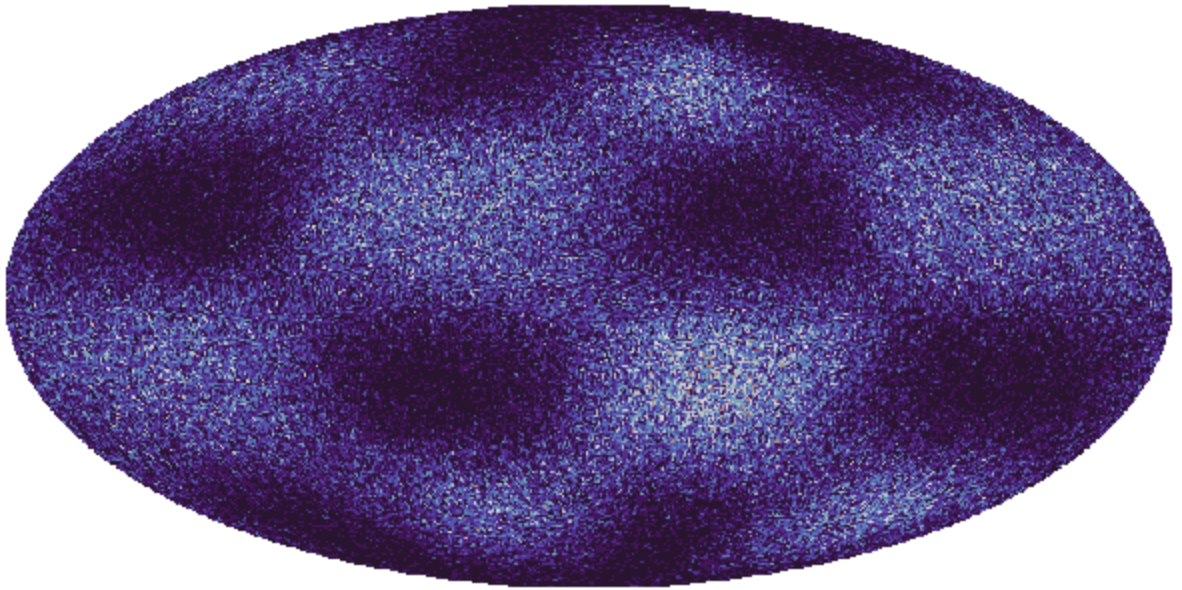} & \includegraphics[width=26mm]{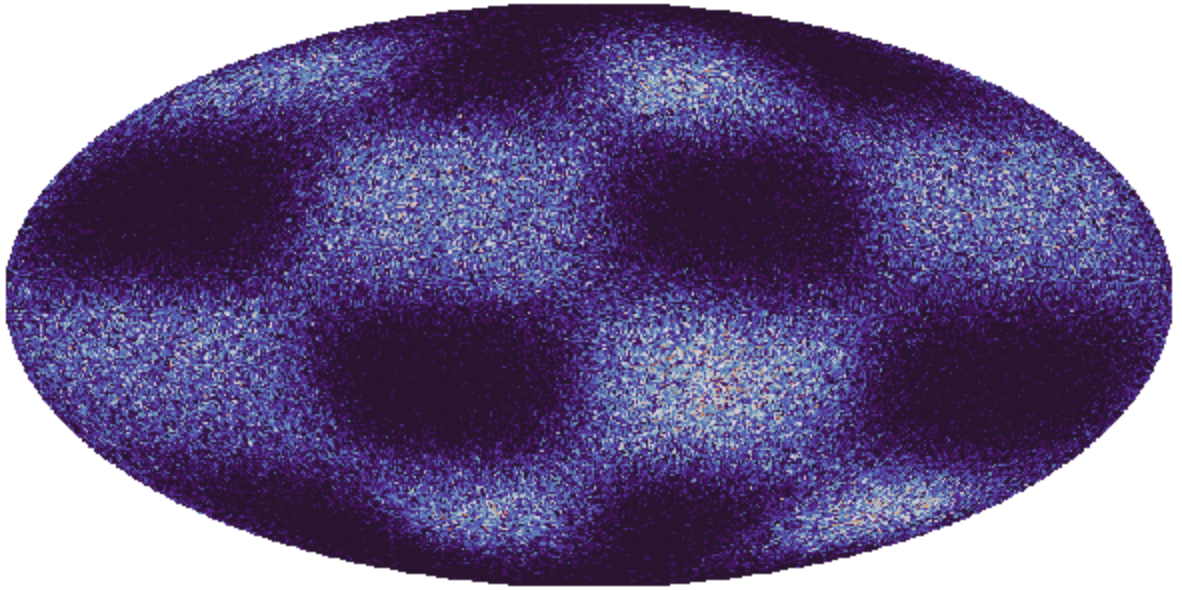} & \includegraphics[width=26mm]{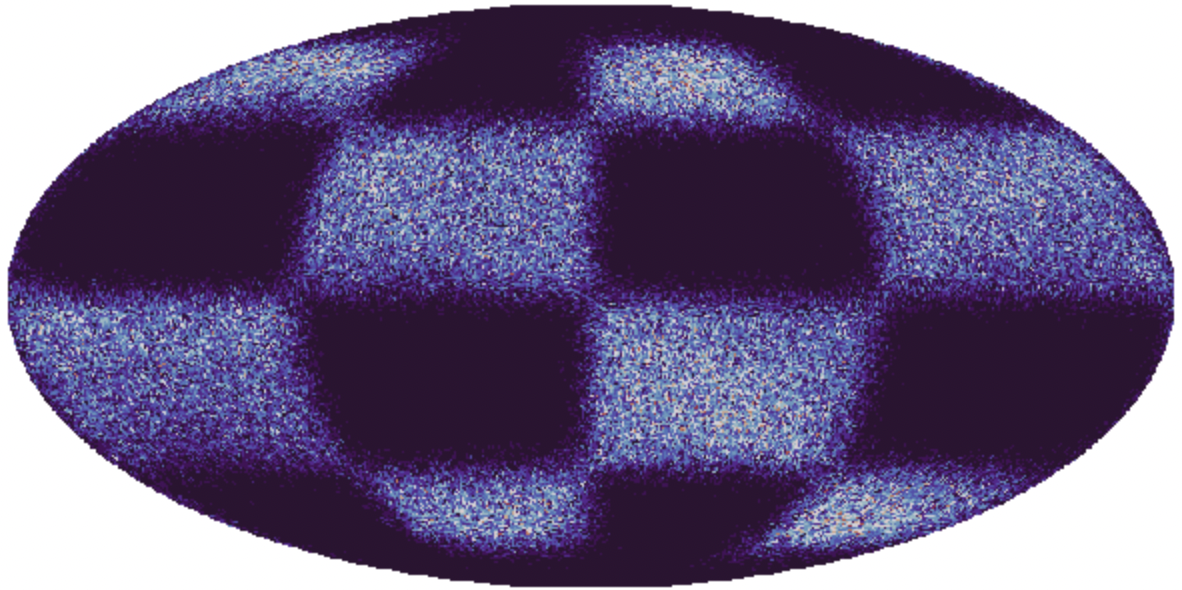}   \\
 
        \end{tabular}
        \caption{\small Learned synthetic density on SO(3). On the left, starting from uniform noise on the sphere at $t=T$, solving the ODE \autoref{eqn:sampling_ode} transports noise samples back into the target density at $t=0$.}
        \label{tbl:toy_density}
    \end{figure}

\section{Application: Emulating Galaxy Intrinsic Alignments in the Illustris-TNG simulations}
Weak gravitational lensing occurs when light rays from distant galaxies get deflected due to the presence of massive objects along their trajectory \citep[e.g.,][]{2001PhR...340..291B}. 
By measuring the coherent shape distortions of ensembles of galaxies, we can study the lensing effect caused by the distribution of matter in the Universe, and thereby learn about dark energy \citep{2015RPPh...78h6901K}. 
One important systematic to model when measuring lensing is the intrinsic alignments (IA) of galaxy shapes; IAs arise due to galaxies tending to point coherently towards other galaxies due to gravitational tidal effects, which mimics a coherent lensing effect \citep{2015PhR...558....1T}.
For cosmological measurements, IA must be taken into account, which means that realistic models for it must be included in synthetic galaxy catalogs.

\paragraph{Cosmological Simulation}
We will explore the efficacy of our model using the hydrodynamical TNG100-1 run at $z=0$ from the IllustrisTNG simulation suite \citep[for more information, please refer to][]{ tng-bimodal,pillepich2018illustristng, Springel2017illustristng, Naiman2018illustristng, Marinacci2017illustristng,tng-publicdata}. We employ a  stellar mass threshold of $ \log_{10}(M_*/M_\odot) \ge 9 $ for all galaxies, using the stellar mass from  their SUBFIND catalog, and select the central galaxies from each group for our analysis. The corresponding host dark matter halos were used to study halo alignments.

\begin{figure*}\label{ellip}
\includegraphics[height=8.5cm ]{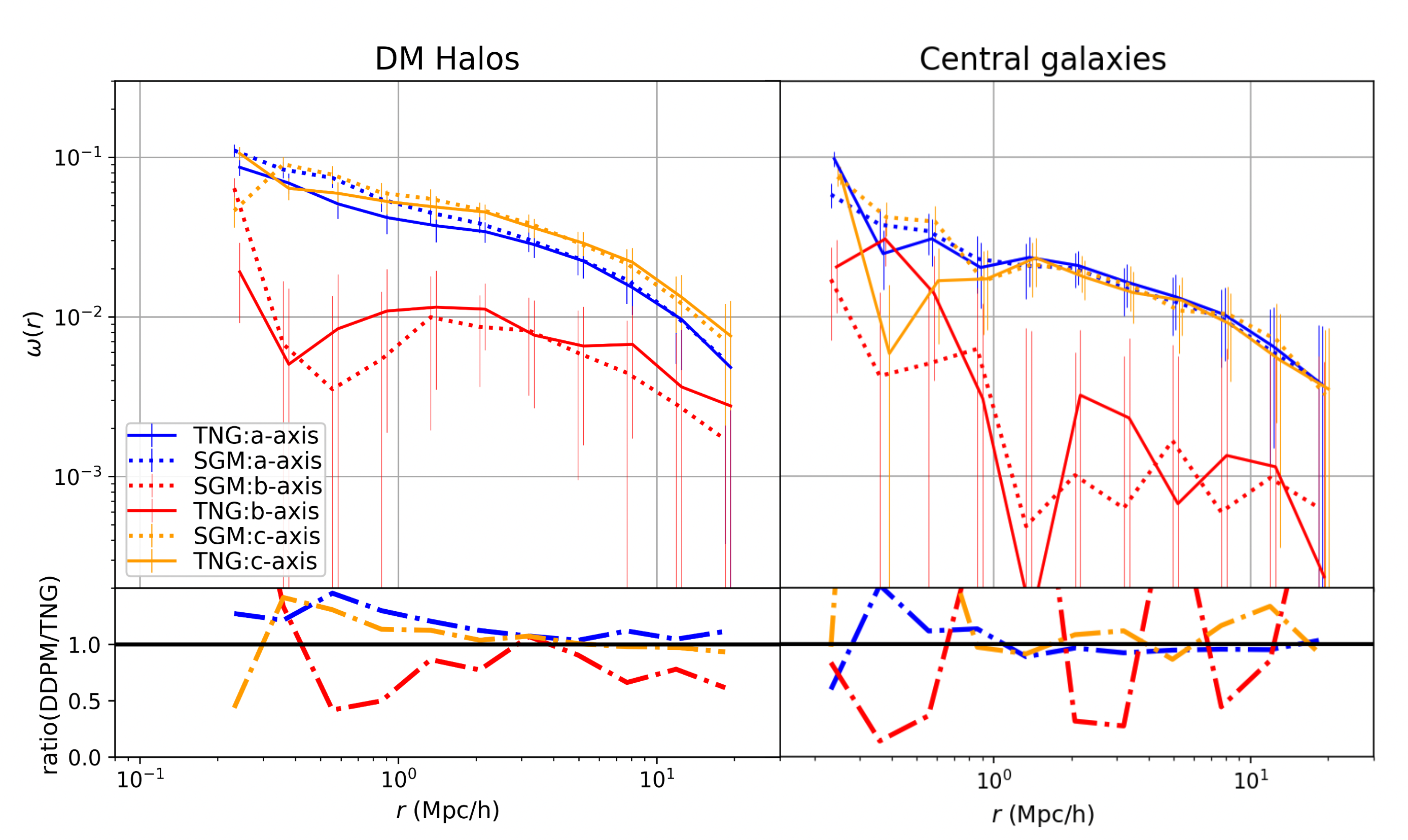}
 
 

\caption{\label{ED_halos}  \small
 The two-point ED correlation function, $\omega(r)$, which captures the correlation between position and the axis direction,
of all galaxy (right) and dark matter halo (left) axes with galaxy positions: the solid lines show the measured values from the TNG simulation, while the dashed lines show the generated values from the SGM. The top panels show  measured $\omega(r)$ values, and the bottom panels show the ratio  $\omega(r)$ from the SGM to that measured in TNG. 
 The SGM curve was shifted by 5 per cent to the left for visual clarity. For the ellipsoid, we denote the major, intermediate, and minor axes as $a$, $b$, and $c$, respectively. }
 \end{figure*}

\paragraph{Results  }\label{results} Throughout the section we  refer to the sample generated from the diffusion model as the \textit{SGM} sample, and the sample from  TNG100 as the \textit{TNG} sample.  
 The inputs to the model are the TNG100 gravitational tidal field (obtained  from the 3D tidal tensor which carries some information about the alignment at large scales), and the outputs are the 3D orientations of halos and central galaxies: the model generates the orientations of halos and galaxies conditioned on the tidal field.
 
 We test our model using the 3D orientation-position correlation function, $\omega(r)$, often referred to as the ED correlation. It captures the correlation between overdensity (galaxy positions) and  orientations of the selected halo/galaxy axes (modeling the halos/galaxies as ellipsoids and selecting either the major, intermediate, or minor axis). Positive  $\omega(r)$ values indicate that the selected halo/galaxy axis exhibits a coherent alignment towards the positions of  nearby galaxies.  The   ED correlation functions for all three axes of the halos and galaxies are presented in Fig.~\ref{ED_halos}.  Here, the errorbars were calculated using the jackknife estimator. 
 In general, the qualitative trend of ED as a function of 3D separation is captured by the SGM for both DM halos and central galaxies. For small  scales (below $r \leq 1 $ Mpc/h), there is a general deviation from the measured values, which may be explained by the highly complex non-linear processes that might not have been captured by the neural network. 
  Quantitatively, for the major axes of both halos and central galaxies, the generated samples agree well with the simulation.
   For the intermediate axes of DM halos and central galaxies, the signal is very weak, though the SGM managed to captured the correlation with statistical consistency. However, for the minor axes, the SGM model slightly underestimates the correlation and overestimates it for central galaxies at small scales.

    Overall, the SGM model can describe synthetic densities with high statistical correlations (as illustrated in \autoref{tbl:toy_density}), and those with low statistical correlations, as shown in the case of galaxy/halo alignments. Regarding the limitations, the model did not capture the correlations at small scales to a good quantitative agreement, for which adding a graph-based layer may help \citep{graphgan}.

\section{Conclusions}\label{conc}
 
 We have introduced a novel score-based generative model for the SO(3) manifold, and applied it in an astrophysical context to the modeling of the 3D orientation of galaxies and dark matter halos in the TNG100 hydrodynamical simulations. 
Predicting galaxy properties given a dark matter halo, or vice versa, is known as the galaxy-halo connection. Deep generative models show promise in tackling this high-dimensional multivariate problem. We have demonstrated that a smaller subset of the problem of modeling halo/galaxy orientation given the tidal field can be addressed with score-based denoising diffusion models. The diffusion model generates orientations that have statistical correlations consistent with those of the cosmological simulation, in addition to reproducing high-correlation synthetic densities on SO(3).
 In the future, we would like to extend this work by implementing a graph layer in order to fully capture the correlation at non-linear (small) scales and extend the number of halo and galaxy properties predicted by the model. Applying our model to a large volume cosmological simulation, to test the ability to model these alignments, will be highly useful for future weak lensing surveys.

\section*{ Broader Impact}\label{bi}
The proposed methodology of deep manifold learning on SO(3) will be practical in many disciplines outside of astrophysics/cosmology. For instance, in robotics the problem of estimating poses of objects is an intensely studied problem and our method provides a way of tackling this problem from a generative perspective with diffusion models. Additionally, in biochemistry it is often hard to find the optimal angle for molecular docking; with our proposed method biochemists could efficiently find the optimal angle that minimizes the potential energy.  
We do not believe that our work poses any negative societal impacts or ethics-related issues. 

\section*{Acknowledgements}
 This work was supported in part by a grant from the Simons Foundation (Simons Investigator in Astrophysics, Award ID 620789) and by the NSF AI Institute: Physics of the Future, NSF PHY- 2020295.

\bibliography{main, references}
\bibliographystyle{plainnat}

 \end{document}